\documentclass[12pt,side,epsfig,citesort,color]{article}
\usepackage{booktabs}
\usepackage{setspace}
\usepackage{color}
\usepackage{geometry}
\usepackage{chemformula}
\usepackage{epsfig}
\usepackage{amssymb}
\usepackage{amsmath}
\usepackage{amsfonts}
\usepackage{graphicx}
\usepackage{epsfig}
\usepackage{braket}
\usepackage{cite}
\usepackage{subfigure}
\usepackage{authblk}

\begin{document}
\voffset=0.9cm

\title{                                                               
\vspace*{-3.2cm}
Interparticle correlations and chemical bonding from physical side: Covalency \textit{versus} atomicity and ionicity }
\author[1]{Ewa Broc\l{}awik} 
\affil[1]{\small{Polish Academy of Arts and Sciences, ul. S\l{}awkowska 17, PL 31-017, Krak\'{o}w, Poland}}
\author[2]{Maciej Fidrysiak} 
\author[2]{Maciej Hendzel} 
\author[2]{J\'{o}zef Spa\l{}ek\thanks{corresponding author: jozef.spalek@uj.edu.pl}
\affil[2]{\small{Institute of Theoretical Physics, Jagiellonian University, ul. \L{}ojasiewicza 11, PL-30-348 Krak\'{o}w, Poland}}                                 
}

\maketitle

\begin{abstract}
In this Chapter we reexamine the concept of \textit{covalency} and \textit{ionicity} on example of the simplest molecules. First, starting from the exact expression for the two--particle wave function in the case of \ch{H2} molecule within the Heitler--London model, we demonstrate an unphysical behavior of the covalency at large interatomic distance which, within standard definition, reaches the maximal value in the limit of separated atoms. Second, we correct this deficiency by introducing the concept of \textit{atomicity}, with the help of which, we define the \textit{true} (intrinsic) \textit{covalency}, as well as retain the precise concept of \textit{ionicity}. We connect the introduced atomicity to the onset of Mott--Hubbard localization, adopted here from the well established notion in the condensed matter. The evolution from the molecular to atomic states develops rapidly with interatomic distance beyond the localization threshold. This brief overview is intented as pedagogical in nature, nonetheless is analyzed quantitatively for the case of \ch{H2} molecule. At the end, we outline a similar--type of model for the related case of the hydrogen bond on example of adenine--tymine pair. Methodologically, the approach is based on combining the first-- (wave mechanics) and second--quantization into a single scheme of formal description.
\end{abstract}

\newpage
\tableofcontents

  \section{Motivation}

The quantum--mechanical concept of the chemical bond was introduced in 
the quantitative way on example of \ch{H2} by Heitler and London (HL) 
(\cite{Heitler--London} see also related papers 
\cite{CondonPNAS, Pauling1928, Coulson1949}). This theory was formulated only a year after the wave 
mechanics had been rigorously established by Schr\"odinger for the 
discrete states of single hydrogen atom and subsequently extended by Dirac \cite{Dirac1926} by introducing the indistinguishability principle to the many--particle wave--function. The HL 
approach was based on what is now known as the
Hartree--Fock two--particle wave function (strictly speaking, solely on its orbital part). This wave function is at present taken of the 
following spin--singlet form 

\begin{align}
    \psi_{12}(\textbf{r}, \textbf{r}) = \frac{1}{\sqrt{2(1+S^2)}} [\varphi_1(\textbf{r}_1)\varphi_2(\textbf{r}_2)+\varphi_1(\textbf{r}_2)\varphi_2(\textbf{r}_1)][\chi_{\uparrow}(1)\chi_{\downarrow}(2)-
    \chi_{\downarrow}(1)\chi_{\uparrow}(2)].
    \label{eq:HLwavefunction}
\end{align}

\noindent
The wave function originally contained hydrogen atomic wave functions 
$\varphi_1(\textbf{r}) = \varphi(\textbf{r}-\textbf{R}_1)$ and 
$\varphi_2(\textbf{r}) = \varphi(\textbf{r}-\textbf{R}_2)$ centered at \ch{H2}
nuclei positions $\textbf{R}_1$ and $\textbf{R}_2$, respectively. They 
represent the hydrogen $1s$ orbital wave functions, whereas 
$\chi_{\sigma}(i)$ are the corresponding spin functions with spin 
$\sigma=\uparrow, \downarrow$ and $i \equiv \textbf{R}_i$. This form \eqref{eq:HLwavefunction} is a result of an {\it ad hoc} assumption and expresses the principle of 
antisymmetry with respect to transposition of the quantum numbers ($i$, 
$\sigma$); $S=\braket{\varphi_1|\varphi_2}$ is the overlap of the functions. 
In general, the evolution of the wave--function form in time is schematically 
summarized in Fig. \ref{tab:wav} (see Table Caption for detailed description)
The ansatz taken to define the form \eqref{eq:HLwavefunction} was later extended by selecting more involved wave functions 
and has resulted in the Full Configuration Interaction (FCI) formulation \cite{Sherill}, involving also
the most important virtually excited states. Nowadays, it comprises a 
whole discipline of advanced computational chemistry, 
including also the Density Functional Theory (DFT) calculations, etc. \cite{Becke}. It 
is this stage at which the unique calculations of Ko\l{}os and 
Wolniewicz resulted in practically exact results for the \ch{H2} molecule ground state
\cite{Kolos1, Kolos2}. Such a procedure also requires, as a must, replacement of the original atomic wave functions 
$\{\varphi_i\}$ in \eqref{eq:HLwavefunction} by 
appropriate molecular orbitals $\{\Phi_i\}$ \cite{Coulson1949}. 

Our approach begins from a different starting point \cite{Hendzel1, Hendzel2, Hendzel3}. Namely, we combine the 
wave ($1^{st}$ quantization) and particle ($2^{nd}$ quantization) 
languages of electron states description, composing a single chemical bond (i.e., of two--electron states) and show how the {\it 
ionicity} arises in a natural manner and, what is crucial, the importance of the introduced concept of 
{\it atomicity} in the bonding state, so that the {\it true} (or {\it
intrinsic}) {\it covalency} can be quantitatively defined. In this manner, an 
intrinsic inconsistency in the standard definition of covalency is 
removed. Furthermore, the full Hamiltonian with all
two--particle interactions can be written in a compact form, as well as 
an accurate analytic expression for the two--particle wave function is explicitly 
provided. This formulation represents the exact solution of the Heitler--London model and constitutes a crucial step 
forward in a systematic analysis of the bonding properties of more complex molecular systems. In particular, the limiting 
situation of separate atoms is recovered correctly, i.e., when the state of indistinguishable (bound) electrons transform 
into that of their distinguishable (atomic) correspondents.  Some of the possible 
extensions of our approach, discussed here in detail for \ch{H2} molecule, are briefly characterized at the end. 
  
\section{Method: First and second quantization combined}
 
The principal features of our original \emph{\textbf{E}xact \textbf{D}iagonalization \textbf{Ab I}nitio} (EDABI) method and interpretation of physical results are carried out here on example of \ch{H2} molecule. Thus the starting point are the electronic states of the two electrons that originate from parent hydrogen atoms, each placed in $1s$ atomic Slater state. We neglect a possibility of virtual electronic transitions to $2s$ and higher excited states and would like to study an adiabatic evolution of those atomic states into the two--particle molecular state (forming the bond) with the decreasing interatomic distance $R$. In the particle (second--quantization) language, the situation can be described by starting from the
field--operator language in which the particles are represented by the field operator of the form 

\begin{align}
    \hat{\psi}_{\sigma}(\textbf{r}) = w_1(\textbf{r})\chi_{\sigma}(1)\hat{a}_{1\sigma} +
    w_2(\textbf{r})\chi_{\sigma}(2)\hat{a}_{2\sigma}
    \label{eq:field_operator}
\end{align}

\noindent
in which the molecular (Wannier) orthogonalized and normalized wave functions $\{w_{i\sigma}(\textbf{r})\equiv w_{\sigma}(\textbf{r}-\textbf{R}_i)\}_{i=1,2}$ are defined as

\begin{align}
    w_{i\sigma}(\textbf{r}) = \beta[\varphi_{i\sigma}(\textbf{r})-\gamma\varphi_{j\sigma}(\textbf{r})] =
    \beta[\phi_{i}(\textbf{r})\chi_{\sigma}(i)-\gamma\phi_{j}(\textbf{r})\chi_{\sigma}(j)],
    \label{wannier}
\end{align}

\noindent
where $\beta$ and $\gamma$ are mixing coefficients for the state centered on site $i$ with that centered on the neighboring
site $j$. We should stress that the selection of the hybridized (Wannier) basis is a natural choice here as only then the 
anticommutators between the creation and annihilation operators have a universal form (see below). 

For the starting atomic states we select Slater--type orbitals 
$\phi_{i\sigma}=\sqrt{\alpha^3/\pi}exp[-\alpha|\textbf{r}-\textbf{R}_i|]\chi_{\sigma}$, in which $\alpha^{-1}$ is an 
adjustable in correlated state size of the orbital (see below). The explicit form of the mixing coefficient is

\begin{align}
    \beta = \frac{1}{\sqrt{2}}\left[\frac{1}{1-S^2}+\frac{1}{\sqrt{1-S^2}}\right]
\end{align}

\noindent
where $\gamma^2 \equiv (1-\beta^2)/\beta$. 

One methodological remark is in place here. Namely, the form 
\eqref{eq:field_operator} of the field operator is approximate, as the 
sum over states $\{w_i(\textbf{r})\}$ should, in principle cover 
\textbf{all} higher excited states, i.e., $\{\phi_i(\textbf{r})\}$ 
should represent a complete set of hydrogen atomic states or the corresponding
full basis of molecular states. For the sake of argument 
clarity the basis is limited to \eqref{eq:field_operator}. However, to 
optimize the truncated energy of the interacting (correlated) 
two--particle state we minimize the ground state energy $E_G$ additionally with 
respect to $\alpha$. In this way, the orbital size is readjusted in the 
resultant interacting (correlated) state. Note that incompleteness of the basis $\{w_i(\textbf{r})\}$
in defining the field operator \eqref{eq:field_operator} means that we reduce here the problem to the 
Heitler--London type of model, albeit we amend it with the single--particle
basis optimization (through adjustment of $\alpha$ in the resultant state) and reanalyse the 
exact solution of the model in the second quantization representation (i.e., going systematically beyond the 
original Hartree--Fock--type solution).

For the selected starting single--particle orthogonalized basis $\{w_{\sigma}(\textbf{r})\}$ we construct the system Hamiltonian in the second--quantized representation which is

\begin{align}
    \hat{\mathcal{H}} = \sum_{\sigma} \int d^3\textbf{r} 
    \hat{\psi}^{\dagger}_{\sigma}(\textbf{r})[-\frac{\hbar}{2m}\nabla^2+V(\textbf{r})]\hat{\psi}_{\sigma}(\textbf{r}) +
    \frac{1}{2}\sum_{\sigma{\sigma}'}\int d^3\textbf{r}d^3{\textbf{r}}'
    \hat{\psi}^{\dagger}_{\sigma}(\textbf{r})\hat{\psi}^{\dagger}_{{\sigma}'}({\textbf{r}}')\frac{e^2}{|\textbf{r}-{\textbf{r}}'|}
    \hat{\psi}_{{\sigma}'}(\textbf{r})\hat{\psi}_{{\sigma}}({\textbf{r}}')
\end{align}

\noindent
where the first term represents the single--particle part containing both kinetic energy of electron and the original atomic potential $V(\textbf{r})$ coming from both protons.  The second term expresses two--particle Coulomb repulsion. Upon inserting expression \eqref{eq:field_operator} for $\hat{\psi}_{\sigma}(\textbf{r})$ and the corresponding one for $\hat{\psi}^{\dagger}_{\sigma}(\textbf{r}) \equiv [\hat{\psi}_{\sigma}(\textbf{r})]^{\dagger}$ we obtain the full Hamiltonian in terms of creation $(\hat{a}^{\dagger}_{i\sigma})$ and annihilation $(\hat{a}_{i\sigma})$ operators, that takes the following form

\begin{align}
     \mathcal{\hat{H}} = & \sum_{i\sigma} \epsilon_i \hat{n}_{i\sigma} 
    + {\sum_{ij\sigma}}' t_{ij}\,\hat{a}^{\dag}_{i\sigma} 
    \,\hat{a}_{j\sigma} +  \sum_{i} U_i 
    \hat{n}_{i\uparrow}\,\hat{n}_{i\downarrow} + \frac{1}{2} 
    {\sum_{ij\sigma{\sigma}'}}' K_{ij}\hat{n}_{i\sigma}\,\hat{n}_{j{\sigma}'}
      \nonumber \\ 
    & - \frac{1}{2}{\sum_{ij}}' J^H_{ij}  \left(\hat{\textbf{S}}_i \cdot
    \hat{\textbf{S}}_j-\frac{1}{4}
    \hat{n}_i\hat{n}_j\right) + \frac{1}{2} {\sum_{ij}}' {J}'_{ij}
    (\hat{a}^{\dagger}_{i\uparrow}\hat{a}^{\dagger}_{i\downarrow}\hat{a
    }_{j\downarrow}\hat{a}_{j\uparrow} + \mathrm{H.c.})  \nonumber \\
    &+ \frac{1}{2} {\sum_{ij}}' V_{ij} 
    (\hat{n}_{i\sigma}+\hat{n}_{j\sigma})(\hat{a}^{\dagger}_{i\bar{\sigma}}\hat{a}_{j\bar{\sigma}}+ \mathrm{H.c.}) + 
    \mathcal{H}_{\text{ion-ion}},
    \label{eq:full_Hamiltonian}
\end{align}

\noindent
In this expression the first term expresses the atomic energy of electrons, whereas $\hat{n}_{i\sigma} \equiv \hat{a}^{\dagger}_{i\sigma}\hat{a}_{i\sigma}$ and $\hat{\textbf{S}}_i \equiv (\hat{a}^{\dagger}_{i\uparrow}\hat{a}_{i\downarrow}, \hat{a}^{\dagger}_{i\downarrow}\hat{a}_{i\uparrow}, \frac{1}{2}(\hat{n}_{i\uparrow}-\hat{n}_{i\downarrow}))$ are the particle--number-- and spin--operators, respectively. The second term represents the hopping energy of electrons between the sites (with $ i \neq j$); it expresses the energy contribution coming from resonant hops of individual carriers between the sites (prime means $i\neq j$). The third term represents the intraatomic part of the Coulomb repulsion between the electrons on the same atom with opposite spins, whereas the fourth contains the corresponding part of the intersite Coulomb (Hubbard) interaction. The next three terms represent direct (Heisenberg) exchange interaction, pair--particle hopping between the sites and the so--called correlated hopping, respectively. The last three terms are usually of lesser importance in correlated systems. The microscopic parameters are integrals containing the Slater single--particle Wannier functions in a standard manner (cf. Appendix in \cite{Spalek2000}) Finally, $\mathcal{H}_{\text{ion--ion}}$ expresses the classical Coulomb interaction between the nuclei (protons in the case of \ch{H2} molecule). Note also that the microscopic parameters $\epsilon_a$, $t_{ij}$, $U$, $K_{ij}$, $J^H_{ij}$, ${J}'_{ij}$, and $V_{ij}$ can be calculated analytically for $1s$ orbitals, as shown by Slater (\cite{Slater}, see also \cite{Hendzel1}); they will be evaluated explicitly here in the interacting (correlated) state (for i.e., readjusted orbital--size)

The next step in process of solving the Hamiltonian is the diagonalization of \eqref{eq:full_Hamiltonian} in the Fock space for given of microscopic parameters. For that purpose, we select a set of trial orthogonal and normalized states for two electrons which in this case of \ch{H2} molecule are

\begin{align}
\begin{cases}
    &\ket{1} = \hat{a}^{\dagger}_{1\uparrow}\hat{a}^{\dagger}_{2\uparrow} \ket{0},\\
    &\ket{2} = \hat{a}^{\dagger}_{1\downarrow}\hat{a}^{\dagger}_{2\downarrow}\ket{0},\\
    &\ket{3} = \frac{1}{\sqrt{2}}(
    \hat{a}^{\dagger}_{1\uparrow}\hat{a}^{\dagger}_{2\downarrow}+\hat{a}^{\dagger}_{1\downarrow}\hat{a}^{\dagger}_{2\uparrow})\ket{0},\\
    &\ket{4} = \frac{1}{\sqrt{2}}(
    \hat{a}^{\dagger}_{1\uparrow}\hat{a}^{\dagger}_{2\downarrow}-\hat{a}^{\dagger}_{1\downarrow}\hat{a}^{\dagger}_{2\uparrow})\ket{0},\\
    &\ket{5} = \frac{1}{\sqrt{2}}(
    \hat{a}^{\dagger}_{1\uparrow}\hat{a}^{\dagger}_{1\downarrow}+\hat{a}^{\dagger}_{2\downarrow}\hat{a}^{\dagger}_{2\uparrow})\ket{0},\\
    &\ket{6} = \frac{1}{\sqrt{2}}(
    \hat{a}^{\dagger}_{1\uparrow}\hat{a}^{\dagger}_{1\downarrow}-\hat{a}^{\dagger}_{2\downarrow}\hat{a}^{\dagger}_{2\uparrow})\ket{0}.
    \end{cases}
    \label{eq:states}
\end{align}

\noindent
Note that for $N$ sites with $N_e$ electrons, we have $\binom{2N}{N_e}$ two--particle states, which is equal to six for $N=2$ sites and $N_e=2$ electrons. The diagonalized state thus contains three spin-triplet states (with the total--spin $S^Z$ component equal to $S^Z=1,-1$, and $0$, respectively) and three spin--singlet states. The basis is complete for this model with two $1s$--type states (HL model) and for Hamiltonian \eqref{eq:full_Hamiltonian} can be brought to the $6\times6$ matrix form, which splits into three $1\times1$ irreducible parts representing three separate spin--triplet states with eigenenergy 

\begin{align}
    \lambda_1 =\lambda_2 = \lambda_3 = \epsilon_1 + \epsilon_2 + K - J^H, 
\end{align}

\noindent 
as well as $3\times3$ submatrix of mixed spin--singlet states, which is of the form
\begin{align}
\hat{\mathcal{H}} =
    \begin{pmatrix}
    \epsilon + K + J^H & 2(t+V)    & 0 \\
    2(t+V)  & 2\epsilon + J + U   & \frac{1}{2}(U_1-U_2) \\ 
    0   & \frac{1}{2}(U_1-U_2) & 2\epsilon + U - J^H
    \end{pmatrix},
    \label{eq:Hammatrix}
\end{align}

\noindent
with $\epsilon_a \equiv (\epsilon_1+\epsilon_2)/2$ and $U \equiv (U_1+U_2)/2$. In the case of two identical atoms (\ch{H2} molecule) this matrix can be diagonalized analytically, (since $U_1 = U_2 = U$),  with the eigenvalues $\lambda_1 = \lambda_2 = \lambda_3 = 2\epsilon_a + K - J$ for the states $\ket{1} = \ket{\lambda_1}$, $\ket{2} = \ket{\lambda_2}$, and $\ket{3} = \ket{\lambda_3}$ and $\lambda_6 = 2\epsilon_a + U - J$ for the state $\ket{6} = \ket{\lambda_6}$. However, the states $\ket{4}$ and $\ket{5}$ are intermixed and in the case $\epsilon_1 = \epsilon_2$, $U_1 = U_2$ (the case of \ch{H2} molecule) the eigenvalues can be written in an analytic form, namely \cite{SpalPSS}

\begin{align}
    \lambda_{4,5} \equiv \lambda_{\pm} = 2\epsilon_a + \frac{1}{2}(U+K) \pm \frac{1}{2}[(U-K)^2+16(t+V)^2]^{\frac{1}{2}}. 
\end{align}

\begin{figure}
    \centering
    \includegraphics[width=0.5\textwidth]{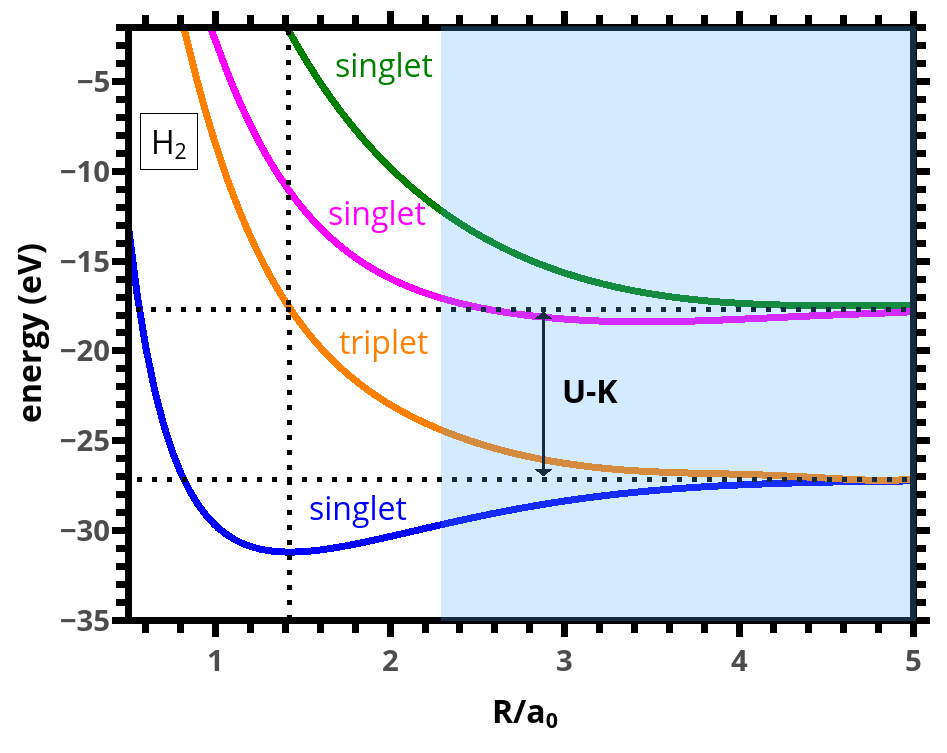}
    \caption{The lowest energy levels composed of three singlet and three triplet states, with the marked Mott regime and associated with it strong-correlation limit (shaded area). The scale $U-K$ represents the effective repulsive Coulomb interaction between electrons, i.e., the HOMO--LUMO splitting. The atomic character of the states shows up gradually becoming atomic with the increasing interatomic distance $R$.}
    \label{fig:el_structure}
\end{figure}

\noindent
The corresponding two eigenstates, in turn, have the following explicit form

\begin{align}
    \ket{\lambda_{4,5}} \equiv \ket{\lambda_{\pm}} = [2D(D\pm U \mp K)]^{-\frac{1}{2}}\{4(t+V)\ket{4}\pm (D \pm U \mp K)\ket{5}\}, 
\end{align}

\noindent
with $D \equiv [(D-K)^2+16(t+V)^2]^{\frac{1}{2}}$. The last two states are the most interesting here, because $\ket{\lambda_{-}}$ is the groundstate; all the remaining states are excited states. We see that the intersite spin singlet states $\ket{\lambda_{\pm}}$ have an admixture of symmetric ionic state $\ket{5}$. This is understood that the bonding $\lambda_{-}$ state for \ch{H2} has an admixture of the ionic states \ch{H+ H-} and \ch{H- H+}. This fact will be interpreted further later. 
\begin{figure}[t]
    \centering
    \includegraphics[width=0.5\textwidth]{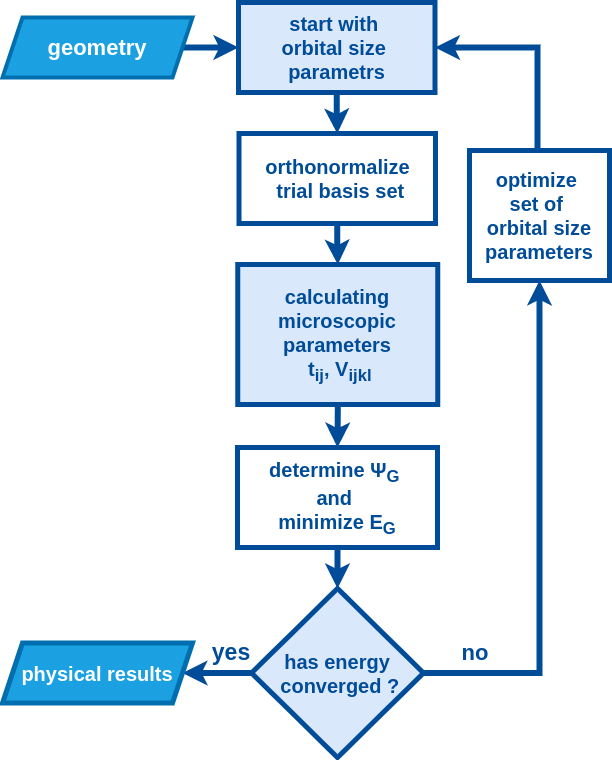}
    \caption{Flowchart of the EDABI method. The method is initialized by selection of a trial single-particle basis of wave functions \eqref{wannier}, and subsequent diagonalization of the many-particle Hamiltonian \eqref{eq:full_Hamiltonian}. Optimization of the single-particle-state size leads to an explicit determination of the trial-wavefunction parameters, microscopic interaction and hopping parameters, as well as ground--state energy, and explicit form of the many-particle wavefunctions (II), all in the correlated interacting state for a given interatomic distance $R$.}
    \label{fig:edabi_scheme}
\end{figure}
To complete the approach we sketch the whole procedure from numerical point of view. First, we 
take the eigenvalue $\lambda_{-}$ as our starting point and minimize this energy with respect to the 
Slater--orbital inverse size $\alpha$, contained in the assumed functions $\{w_i(\textbf{r})\}$. This means that 
we iterate the procedure depicted in Fig. \ref{fig:edabi_scheme} until it converges for optimal $\alpha=\alpha_0$ and for a given interatomic
distance. The optimal value of $\alpha=\alpha_0(R)$ is subsequently the same for all the 
eigenstates $\lambda_1$, $\lambda_2$, $\lambda_3$, $\lambda_4$, $\lambda_5$, and $\lambda_6$. The ground 
state and the next five excited electron energies (for fixed $R = |\textbf{R}|$) are exhibited in Fig. 
\ref{fig:el_structure}. Those results are for \ch{H2} molecule; the other examples, \ch{LiH} and 
\ch{HeH+}, are discussed briefly later. They are of limited absolute accuracy, but these results are only a starting point
to a deeper analysis discussed next. 

\section{True covalency, ionicity, atomicity: \ch{H2} molecule}

\subsection{Two--particle wave function and its basic properties -- analytic solution}

Further analysis of the bonding requires a calculation of the two--particle wave function in the position (Schr\"odinger) representation. For that purpose, we quote a general theorem about the $N$--particle wave function representation in the Fock space (see e.g. \cite{Robertson})

\begin{align}
    \ket{\psi_N} = \frac{1}{\sqrt{N!}} \int d^3\textbf{r}_1d^3\textbf{r}_N \psi_N(\textbf{r}_1,...,\textbf{r}_N)\hat{\psi}^{\dagger}_1(\textbf{r}_1)...\hat{\psi}^{\dagger}_M(\textbf{r}_N)\ket{0}, 
\end{align}

\noindent
where $\psi_N(\textbf{r}_1,...,\textbf{r}_N)$ is the desired wave function for state $N$ in the position representation and $\ket{0}$ represents the vacuum state in the Fock space. We can extract the wave function by reversing the above relation. This leads to the following general explicit expression

\begin{align}
    \psi_{\alpha}(\textbf{r}_1,...,\textbf{r}_N) = \frac{1}{\sqrt{N!}}\bra{0}\hat{\psi}_1(\textbf{r}_1)...\hat{\psi}_N(\textbf{r}_N)\ket{\lambda_{\alpha}}, 
\end{align}

\noindent
where $\ket{\lambda}$ is the eigenstate, for which the wave function $\psi_{\sigma}$ is explicitly determined by replacing $\ket{\psi_N}$ with the corresponding Fock--space eigenstate $\ket{\lambda_{\alpha}}$. For the spin--conserving interaction here, $\sigma \equiv (\sigma_1,...,\sigma_N)$ is fixed $N$--spin configuration. In what follows we discuss the situation for $N=2$ and $\sigma \equiv (\sigma_1,\sigma_2)$ spin--singlet ground state configuration for the \ch{H2} molecule. Namely, the wave function is obtained by utilizing the expression \eqref{eq:field_operator} for the field operators and the anti--commutation relations of the creation and annihilation operators, namely

\begin{align}
    \begin{cases}
        \hat{a}_{i\sigma}\hat{a}^{\dagger}_{j\sigma'}+\hat{a}_{j\sigma'}\hat{a}^{\dagger}_{i\sigma} = \delta_{ij}\delta_{\sigma{\sigma}'}, \\
        \hat{a}_{i\sigma}\hat{a}_{j\sigma'}+\hat{a}_{j\sigma'}\hat{a}_{i\sigma} = 0. 
    \end{cases}
\end{align}

\noindent
Additionally, we make use of the vacuum--state property $\hat{a}_{i\sigma}\ket{0} \equiv 0$ for every state $\ket{i}$ (cf. \eqref{eq:states}).

In effect, the two--particle wave function representing a single bond in the ground state has the following explicit form

\begin{align}
     \Psi_0(\textbf{r}_1,\textbf{r}_2) = \frac{2(t+V)}{\sqrt{2D(D-U+K)}} \Psi_{cov}(\textbf{r}_1,\textbf{r}_2)
     -\frac{1}{2}\sqrt{\frac{D-U+K}{2D}}\Psi_{ion}(\textbf{r}_1,\textbf{r}_2), 
     \label{coeff_wav}
\end{align}

\noindent
where the covalent ($\Psi_{cov}$) and ionic ($\Psi_{ion}$) components are 

\begin{align}
    \Psi_{cov}(\textbf{r}_1,\textbf{r}_2) =& \left[ 
    w_1(\textbf{r}_1)w_2(\textbf{r}_2)+w_1(\textbf{r}_2)w_2(\textbf{r}_1)\right] \left[
    \chi_{\uparrow}(\textbf{r}_1)\chi_{\downarrow}(\textbf{r}_2)-\chi_{\downarrow}(\textbf{r}_1)\chi_{\uparrow}(\textbf{r}_2)\right], 
    \label{wav1} \\
    \Psi_{ion}(\textbf{r}_1,\textbf{r}_2) =& \left[  
    w_1(\textbf{r}_1)w_1(\textbf{r}_2)+w_2(\textbf{r}_1)w_2(\textbf{r}_2)\right] \left[  
    \chi_{\uparrow}(\textbf{r}_1)\chi_{\downarrow}(\textbf{r}_2)-\chi_{\downarrow}(\textbf{r}_1)\chi_{\uparrow}(\textbf{r}_2)\right]. 
    \label{wav2}
\end{align}

\noindent
From Eqs. \eqref{wav1} and \eqref{wav2} it is evident that the part \eqref{wav1} represents  the covalent (cov) part, whereas \eqref{wav2} represents the combination of the two ionic (ion) configurations of the electrons. Also, note that those wave functions have an involved form; spatially symmetric and spin antisymmetric, as it should be for the spin--singlet state. Parenthetically, these are not single Slater determinant states. Also, their factorization into space and spin parts reflects the fundamental properly that they represent separate orbital and internal (spin) symmetries (cf. \cite{DiracP}). The coefficients before the term contain all interactions present in the two--particle single--orbital quantum state. The wave function \eqref{coeff_wav} represents the exact wave function for the Heitler--London model, with additional readjustment of orbital size in the resultant (correlated) eigenstate.

In order to interpret the wave function \eqref{coeff_wav} in terms of the original Slater (atomic) wave functions $\psi_i(\textbf{r}) \equiv \psi(\textbf{r}-\textbf{R}_i)$, we make use of transformation \eqref{wannier} and obtain 

\begin{align}
\begin{split}
     &\Psi_0(\textbf{r}_1,\textbf{r}_2) = \left(C\beta^2(1+\gamma^2) - 2\gamma I \beta^2\right)\phi^{at}_{cov}(\textbf{r}_1,\textbf{r}_2)  
     +\left( I\beta^2(1-\gamma^2) - 2\gamma C \beta^2 \right)\phi^{at}_{ion}(\textbf{r}_1,\textbf{r}_2) \\
     &\equiv \tilde{C}\phi^{at}_{cov} + \tilde{I}\phi^{at}_{ion},
     \end{split}
     \label{coeff_wav2}
\end{align}

\begin{figure}
    \centering
    \includegraphics[width=0.8\textwidth]{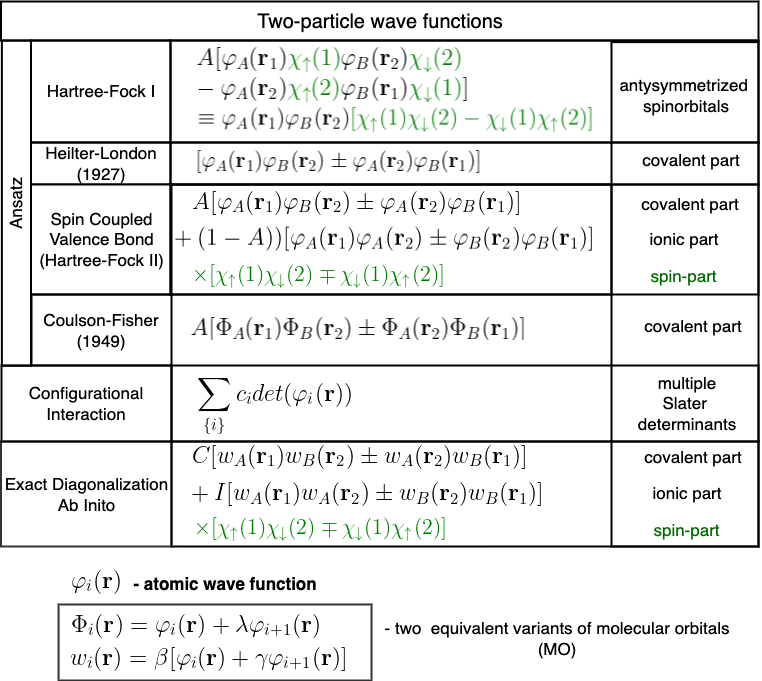}
    \caption{The evolution of the form of the two--particle wave function starting from original (Hartree--Fock I), through Heitler--London (originally without the spin part), followed by spin--coupled valence bond from Coulson and Fisher, as well as the Configuration--Interaction form (the last represents linear combination of Slater determinants for possible occupancies). The bottom line composes our result obtained on the basis of our EDABI method.}
    \label{tab:wav}
\end{figure}
\noindent
where the two--particle functions $\phi^{at}_{cov}(\textbf{r}_1, \textbf{r}_2)$ and $\phi^{at}_{ion}(\textbf{r}_1, \textbf{r}_2)$ have the same formal expressions as \eqref{wav1} and \eqref{wav2}, respectively, but with molecular
functions $\{w_i(\textbf{r})\}$ being replaced by atomic orbitals $\{\varphi_i(\textbf{r})\}$. This last expression coincides with the spin--valence bond wave function, except that the postulated coefficients A and (1-A) before the component functions (cf. Eq. \ref{tab:wav}) are here calculated microscopically. Additionally,
as said above, the Slater--orbital size is optimized in the resultant (interacting) ground state. Therefore, one can say that the presented solution exemplifies a direct and exact solution of the original formulation of Heitler--London model with subsequent optimization of the initial atomic states. More importantly, the intrinsic covalent bonding (see below) will be defined as a contribution calculated beyond that obtained from HL approach. Achieving these goals is the principal aim of the article so far. Note that the solution in its fully analytic form, but requires the utilization of both first-- and second--quantization (wave and particle) aspects of the problem. In the following subsection, we interpret in detail the obtained results and revise the covalency concept. 

\subsection{Towards complementary characterization of the chemical bond: The case of \ch{H2} molecule}

\begin{figure}
    \centering
    \includegraphics[width=0.65\textwidth]{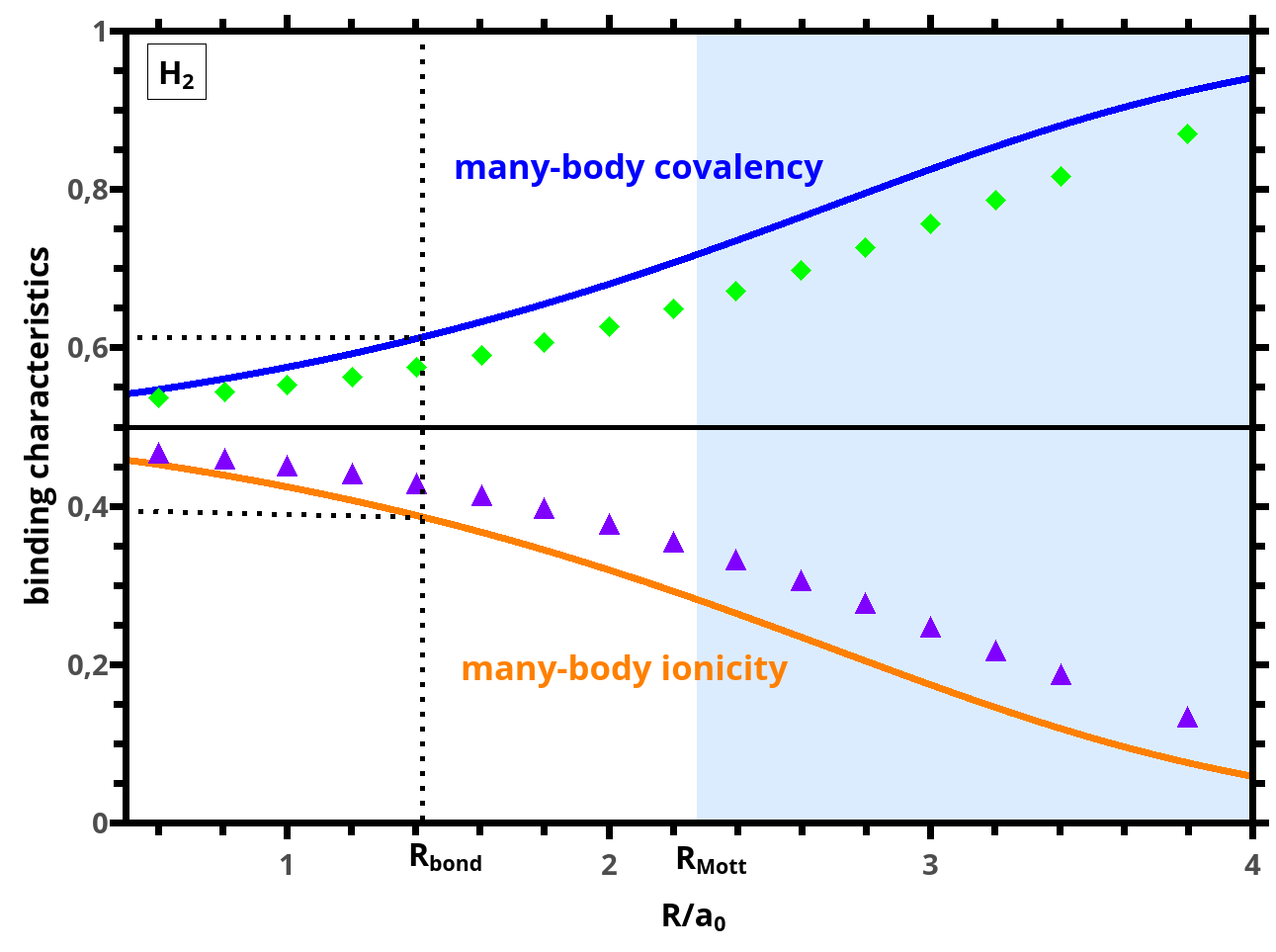}
    \caption{Two-particle covalency vs. corresponding ionicity for \ch{H2} molecule, calculated within EDABI method and compared with the results of Ref.~\cite{pendas}. Shaded regime marks a gradual evolution towards \emph{atomicity}, as determined from the Mott-Hubbard criterion (see in the main text). The vertical dotted line marks equilibrium interatomic distance, whereas the horizontal dotted lines illustrate the dominant character of the covalency in that state (with the ratio $r = 1.43 \sim 2:1$)).}
    \label{fig:cov_wrong}
\end{figure}
The wave function \eqref{coeff_wav2} 
reduces to the Heitler--London form after taking $\tilde{I} \equiv 0$ and neglecting the spin part, as well as disregarding the readjustment the orbital size in the correlated state. But, most importantly, neglecting to mix the Slater orbitals, i.e., putting in Eq. \eqref{wannier} $\beta\equiv 1$, and $\gamma \equiv 0$. Such an assumption leads to the two--particle wave function \eqref{coeff_wav2} in terms of atomic Slater orbitals and without bare covalency (single--particle mixing) \cite{PaulingNature}. This leaves us with the principal question to what extent the factor $\tilde{C}$ in \eqref{coeff_wav2} describes the real covalency, as the $\phi^{at}_{cov}(\textbf{r}_1, \textbf{r}_2)$ in the original Heitler--London approach describes the whole covalency. The fact that it is not the whole story, even when $\gamma \neq 0$, can be demonstrated explicitly by taking $lim_{R\rightarrow \infty}\text{ }\gamma_{cov} = lim_{\gamma\rightarrow 0}\text{ }\gamma_{cov} = 1$. This is a clearly unphysical result. The reason why it is so is coded in choosing the symmetric form of
$\psi^{at}_{cov}(\textbf{r}_1, \textbf{r}_2)$, which amounts to selecting that function in the form for \textit{indistinguishable} particles, whereas in the limit $R \rightarrow \infty$ the electrons are located on separated atoms and hence are clearly \textit{distinguishable} in the quantum--mechanical sense. The unphysical behavior of the covalency factor $\gamma_{cov} \equiv \gamma_{cov}(R)$ is exhibited explicitly in Fig. \ref{fig:cov_wrong}. This behavior is quite astonishing in view of the fact that the same form of postulated (\textit{ad hoc}) wave function \eqref{coeff_wav} provides a good semi--quantitative estimate of the binding
energy and the bond length. In retrospect, the definition of covalency in the Heitler--London fashion may be expected as insufficient, since it contains as factors the pure atomic functions, which in turn, overestimate the atomicity mixed up with the covalency, as elaborated next.
 
In order to remove this inconsistency, appearing practically in every approach of selecting the two--particle form 
based on the indistinguishability of the particles, irrespective of the interatomic distance $\textbf{R}$, we have the following bold proposal. Namely, to define intrinsic or true covalency we extract from the covalency $\gamma_{cov}$ in \eqref{coeff_wav2} (or, equivalently in \eqref{coeff_wav}) the part $\gamma_{cov}$ taken in the limit $\gamma = 0$. This is carried out for each $R$. This step does not mean that we mix up \textit{indistinguishable} and  \textit{distinguishable} states,
as it would be principally incorrect. It merely helps to define the \textit{intrinsic covalency} (and degree of \textit{atomicity}) 
in any nominally covalent system. To summarize, we formally define the true covalency, atomicity, and ionicity for each $R$ as 

\begin{align}
    &\text{covalency: } \gamma_{cov} \equiv \frac{|\tilde{C}|^2-|\tilde{A}|^2}{|\tilde{C}|^2+|\tilde{I}|^2},\label{cov} \\ 
    &\text{ionicity: }\gamma_{ion} \equiv 
    \frac{|\tilde{I}|^2}{|\tilde{C}|^2+|\tilde{I}|^2}, \label{ion}\\ 
    &\text{atomicity: }\gamma_{at} \equiv
    \frac{|\tilde{A}|^2}{|\tilde{C}|^2+|\tilde{I}|^2}. \label{atom}
\end{align}

\noindent
Note that the sum of contributions (two--particle probabilities) is equal to unity. The atomicity is incorporated on the classical level (through the probability).  

Next, we incorporate quantitatively the introduced characteristics into the interpretation scheme, but first relate the atomicity to the Mott--Hubbard localization onset, the latter transferred here from the condensed--matter states and adopted to the molecular (finite--size) systems.

\subsection{Atomicity as the onset of localization and a consistent characterization of the chemical bond}

In general, the atomicity here is purposefully associated with the Mott--Hubbard localization, as it provides a physical rationale behind our concept. The localization on parent atoms of electrons in delocalized (band or other) states takes place under a gradually increasing repulsive interaction among the involved particles. In other words, it represents atomization of particles from collective states in lattice systems experiencing an increasing magnitude of repulsive interaction between them. This can be achieved either by increasing the interatomic distance and/or decreasing particle density in the system. In our case here the atomization can be gradual with the increasing interatomic distance. To elucidate how this happens we relate it to the 
well known Mott \cite{Mott} and Hubbard \cite{Hubbard} criteria for onsets of localized (atomic) behavior of electrons. This is because, we would like to interpret
the evolution of molecular \ch{H2} (electron--paired) state into individual singly occupied (atomic) as a gradual process of the Mott--Hubbard localization. Namely, we define the Hubbard and Mott onset criteria adapted to the present situation as 

\begin{align}
    \frac{2|t+V|}{U-K} = 1, \text{      or      } n^{1/d}_c \alpha^{-1} \equiv \frac{1}{\alpha_0 R_{Mott}} \simeq 0.5, 
    \label{Hubbard-Mott}
\end{align}

\noindent
respectively, where $\alpha^{-1}_0$ is the readjusted orbital size at $R=R_{Mott}$ (note, the criteria define, explicitly $R_{Mott}$).
The first of them means that, for $R=R_{Mott}$, the kinetic (hopping) energy, arising from \eqref{eq:full_Hamiltonian} is equal to the interaction energy. That is, for $R<R_{Mott}$ the ratio is greater than unity, whereas for $R>R_{Mott}$ it is smaller than unity and approaches relatively rapidly zero with $R \rightarrow \infty$ (beyond $R_{Mott}$). Physically, the kinetic energy predominates in the former case; this hopping of electrons, in practice, gets gradually frozen out on atoms as $R$ increases beyond $R_{Mott}$. On the other hand, the Mott criterion expresses the localization onset when the diameter of the renormalized orbital ($2\alpha^{-1}_0$) crosses the value of interatomic distance; this happens at $R\simeq R_{Mott}$. Quasiclassicaly, the latter criterion means that the collective electron behavior 
is established when their orbitals start overlapping strongly.

\begin{figure}[t]
    \centering
    \includegraphics[width=0.6\textwidth]{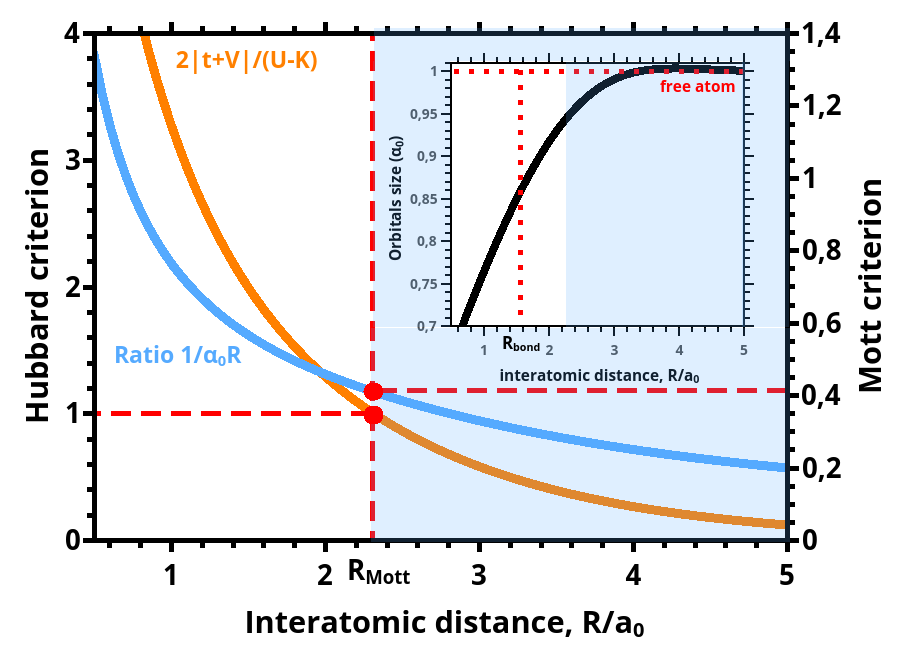}
    \caption{Hubbard (orange) and Mott (blue) characteristics of atomicity vs. interatomic distance $R$ (dashed horizontal lines). The dots mark the points corresponding to Hubbard and Mott criteria. The vertical dotted line marks the onset of \textit{Mottness} at $R_{Mott}$. The inset: $R$ dependence of the orbital size of the renormalized atomic wave functions composing the molecular (Wannier) single-particle states. The dotted line marks the equilibrium distance $R_{bond}$.}
    \label{fig:localization}
\end{figure}
To visualize this physical and formal reasoning we have plotted in Fig. \ref{fig:localization} the left parts of \eqref{Hubbard-Mott} \textit{versus} $R$, as well as have marked explicitly their values at $R=R_{Mott}$. The blue shaded area may be called \textit{the Mott regime}. Additionally, in the inset we display the $R$ dependence of the readjusted orbital size $\alpha^{-1}_0$; it approaches rapidly the atomic value $a_0 = 0.53 \AA$ upon entering the Mott regime. This feature illustrates the rate at which the atomic states are being established with the increasing $R$ beyond $R_{Mott}$.
\begin{figure}[t]
    \centering
    \includegraphics[width=0.6\textwidth]{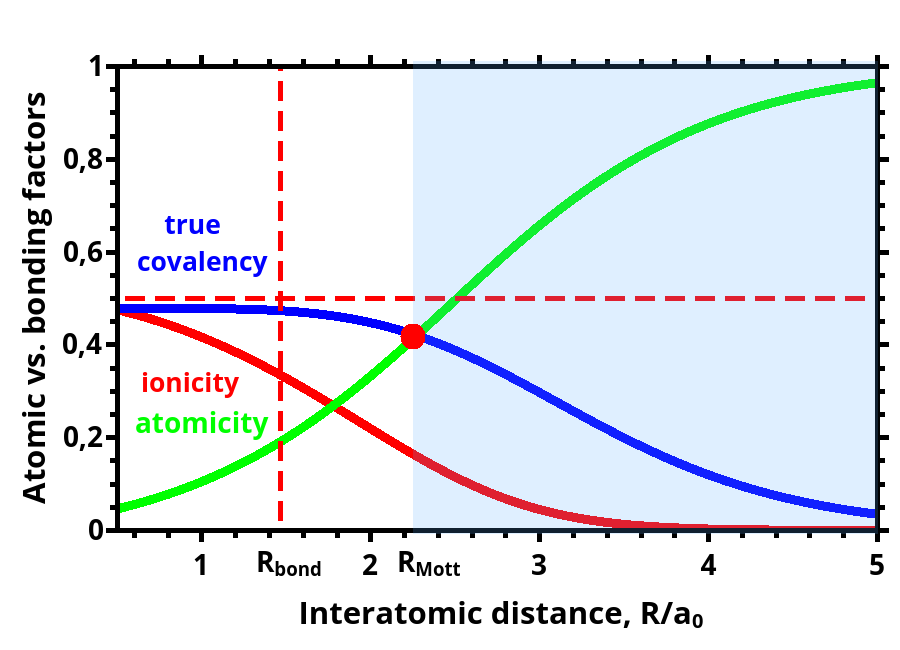}
    \caption{Intrinsic properties of the chemical bond: atomicity (green), true covalency (blue), and ionicity (red), all as a function of interatomic distance $R$. They represent the relative weights in the total two-particle wave function. In the $R\rightarrow 0$ limit the atomicity practically disappears and is the only contribution in the separate--atom limit $R \rightarrow \infty$. The solid circle defines the onset of localization effects (\textit{Mottness}) due to interelectronic correlations. If the atomicity is disregarded, the covalency exhibits a drastic nonphysical behavior with increasing $R > R_{bond}$. The figure illustrates a systematic evolution of molecular states into separate atoms and vice versa, formation of molecular states out of separate atoms. The Slater states have a renormalized size $\alpha^{-1}\leq\alpha_B$.}
    \label{fig:covionat}
\end{figure}

The connection of the above criteria with the true covalency ($\gamma_{cov}$), atomicity ($\gamma_{at}$), and ionicity ($\gamma_{ion}$) is shown explicitly in Fig. \ref{fig:covionat}, where the $R$ dependence of those quantities is drawn. Quite remarkably, at $R=R_{Mott}$ the true covalency and atomicity acquire the same value, what illustrates decidedly that the point $R_{Mott}$ expresses a crossover point from covalency to atomicity--dominated regime. In the 
complementary regime $R\rightarrow 0$, the true covalency and ionicity (atom double occupancy) gradually coalesce. Both figures provide a combined picture in accord with our physical intuition. Simply, the criteria depicted in Figs. \ref{fig:localization} and 
\ref{fig:covionat} are mutually consistent and complementary to each other. As we concluded in \cite{Hendzel2}: \textit{This agreement leads to the conclusion that the introduced entities (19)-(21) and (22) are not only relevant in condensed--matter (extended) systems, but also appear as a crucial incipient feature in molecular systems.} Such characterization is possible only with introducing microscopically derived two--particle wave function \eqref{coeff_wav} (or \eqref{coeff_wav2}) as the proper characteristic of a single bond, which, after all, is composed of electron pair. 

We can illustrate further the onset of partial electron localization by calculating directly the density of electrons in the ground state from the formula the taken from second--quantization scheme, i.e.

\begin{align}
    n_{\sigma}(\textbf{r}) = \braket{\lambda_5|\hat{\psi}^{\dagger}_{\sigma}(\textbf{r})\hat{\psi}_{\sigma}(\textbf{r})|\lambda_5}. \label{n_particle}
\end{align}

\noindent
\begin{figure}[]
    \centering
    \includegraphics[width=1\textwidth]{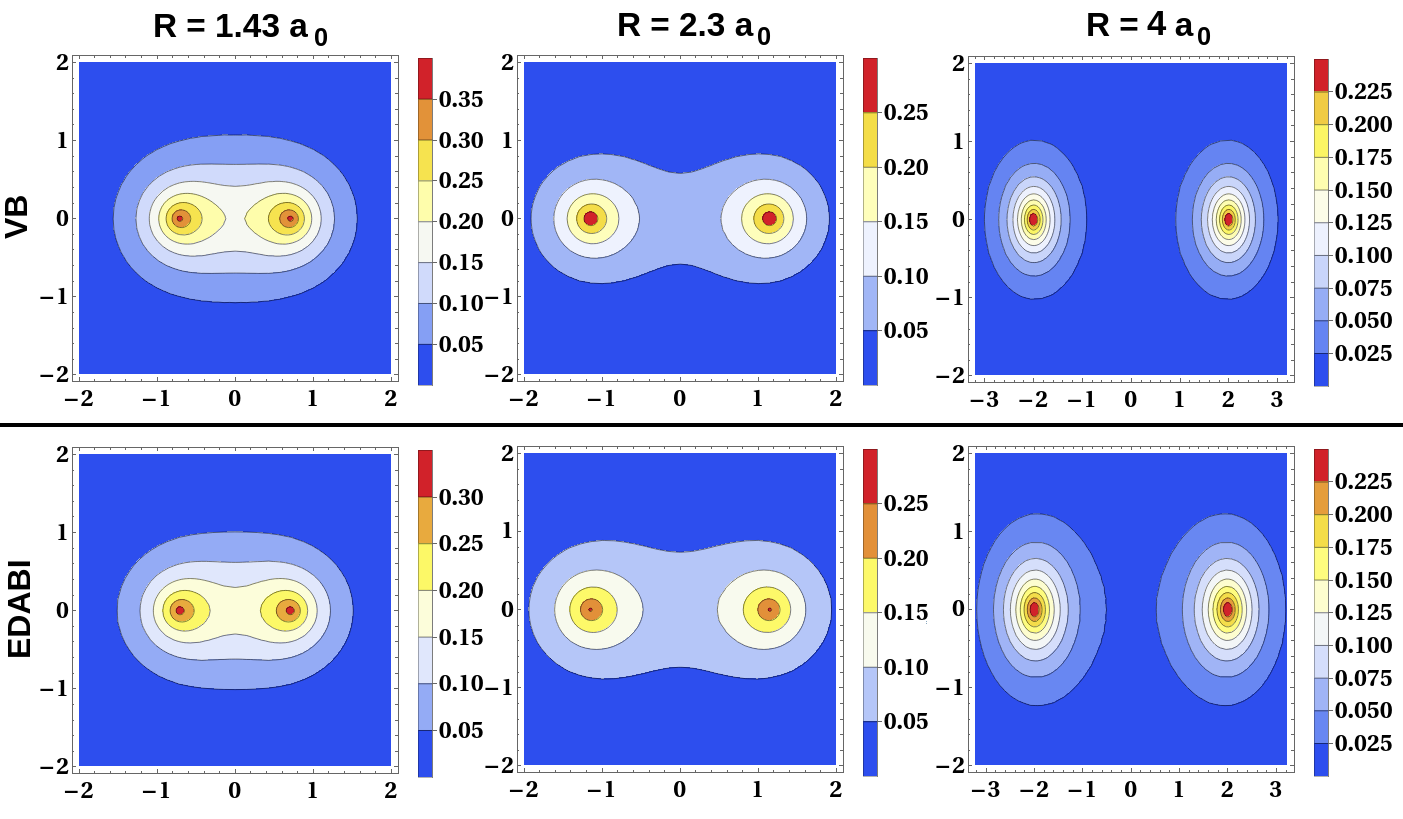}
    \caption{Electron density $n_{\sigma}(\textbf{r})$ for different interatomic distances: $R=1.43a_0^{-1}$, $R=2.3a_0^{-1}$, and $R=4a_0^{-1}$. The parts centered at nuclei are practically disjoint for $R \gtrsim 4a_0$, illustrating the robustness of atomic behavior in that situation. This density contains also the double-occupancy (ionicity) contribution which is becoming rapidly negligible with the increasing $R$ beyond $R_{Mott}$. The evolution towards the atomic states beyond $R_{Mott}$ is more rapid in the EDABI approach, even so the Heitler--London wave function overestimates the atomicity, as it is composed of bare atomic wave functions.}
    \label{fig:density}
\end{figure}
Note that $ n_{\sigma}(\textbf{r})$ represents the electron density with spin $\sigma$; it is related to the total particle density by $n(\textbf{r}) = 2n_{\sigma}(\textbf{r})$ in the spin--singlet state. This density represents complementary quantity to the probability density $|\psi_0(\textbf{r}_1, \textbf{r}_2)|^2$ in the sense that the former represents the physical--particle--occupancy density profile of $n_{\sigma}(\textbf{r})$ that has been displayed in the panel in Fig. \ref{fig:density} for characteristic interatomic distances $R = R_{bond}=1.43 a_0^{-1}$, $R = R_{Mott}=2.3a_0^{-1}$, and $R=4a_0^{-1}$. The parts centered at the nuclei are practically disjoint for $R \gtrsim 4a_0 $. For comparison, the results from the Valence Bond (VB) approach have been also displayed. In both approaches the behavior of the density profiles with the increasing distance is qualitatively similar. The reason is that in both cases the atomicity part has not been excluded. From this point of view, a direct definition \eqref{atom} of atomicity is the only explicit characteristic. 

\section{Many--body covalency in related systems}

\subsection{\ch{LiH} and \ch{HeH+}}
\begin{figure}[htbp]
    \centering
    \includegraphics[width=1\textwidth]{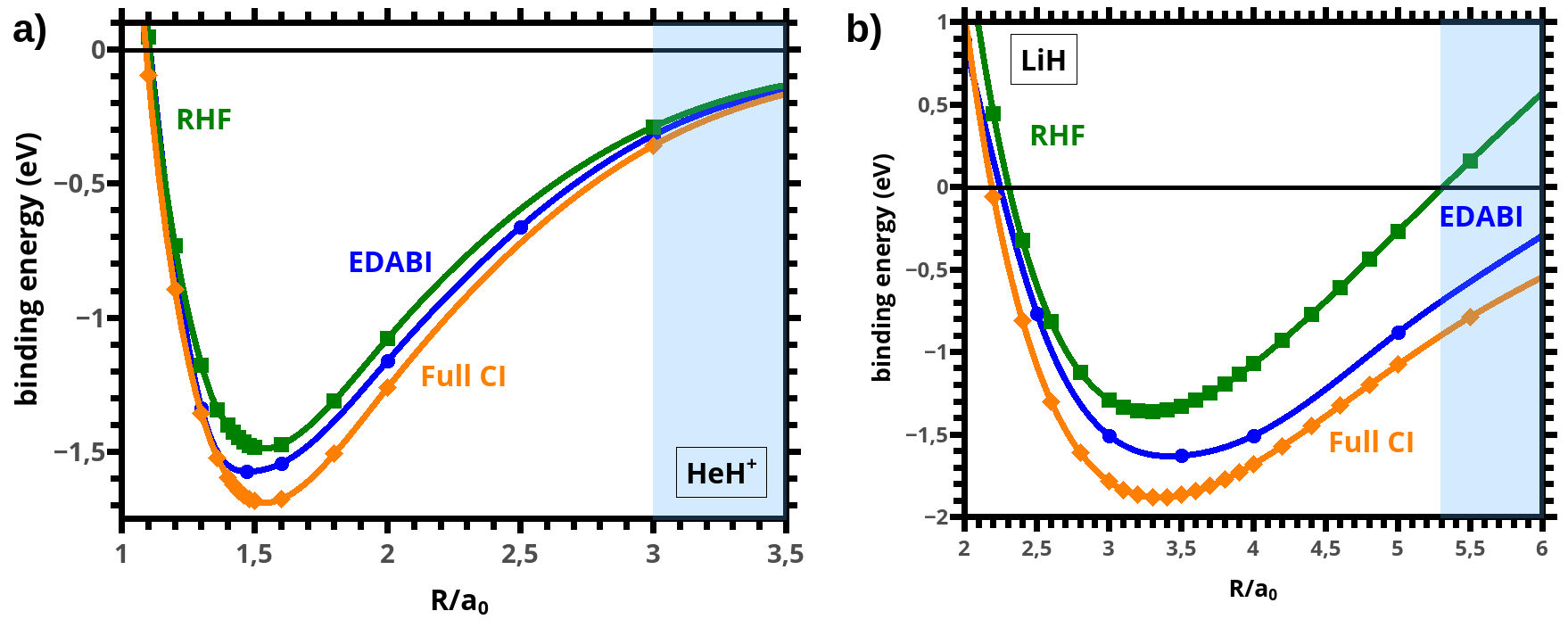}
    \caption{The \ch{HeH+} and \ch{LiH} binding energies versus relative interatomic distance, obtained using EDABI method and compared with restricted Hartree-Fock (RHF) and full configuration interaction (full CI) approach. $a_0 = 0.53\AA$ is the Bohr radius.}
    \label{fig:bind_heh}
\end{figure}

\begin{figure}[htbp]
    \centering
    \includegraphics[width=1\textwidth]{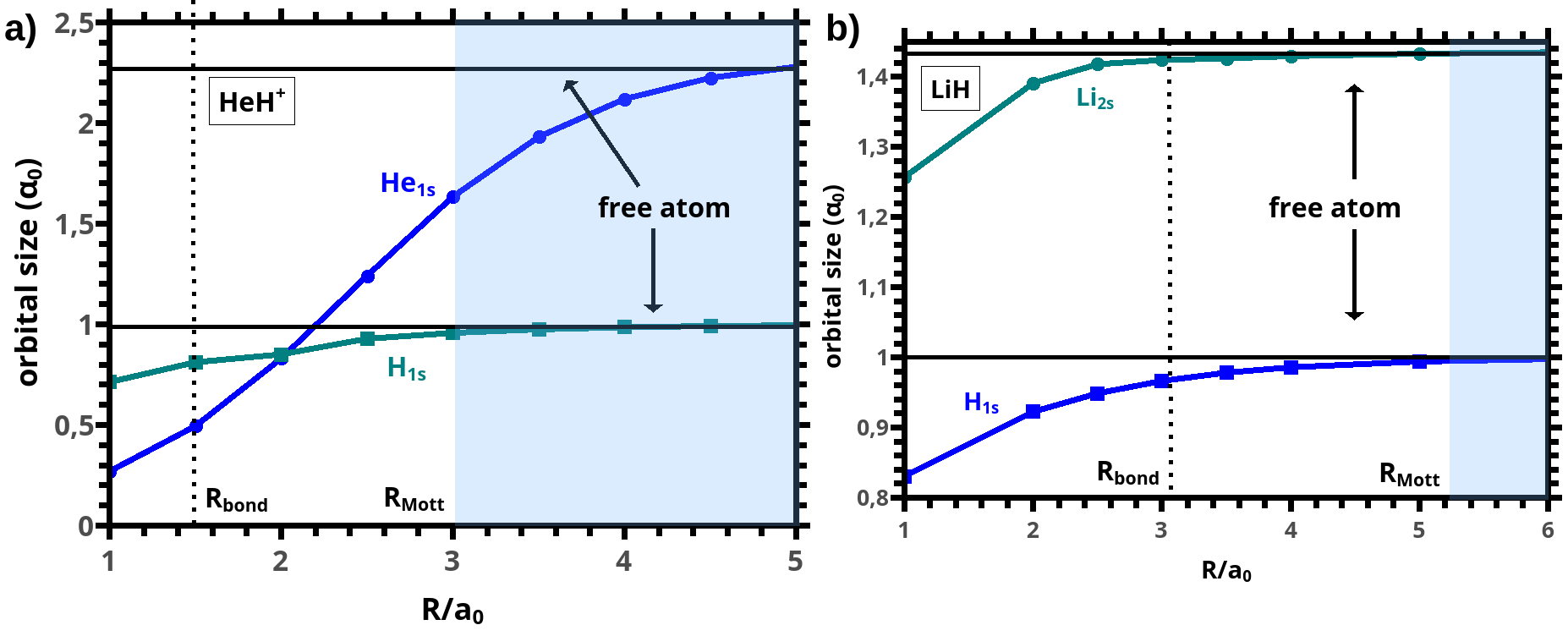}
    \caption{Renormalized $1s$ and $2s$ orbitals size $\alpha^{-1}$ (in Bohr units $a_0$) vs. relative interatomic distance for the $\ch{H2}$ molecule. Note that after crossing the Mott-Hubbard point $R=R_{\mathrm{Mott}}$, $\alpha^{-1}$ approaches rapidly its atomic-limit value $\alpha^{-1}=a_0$. }
    \label{fig:alpha_h2}
\end{figure}

\begin{figure}{}
    \centering
    \includegraphics[width=0.6\textwidth]{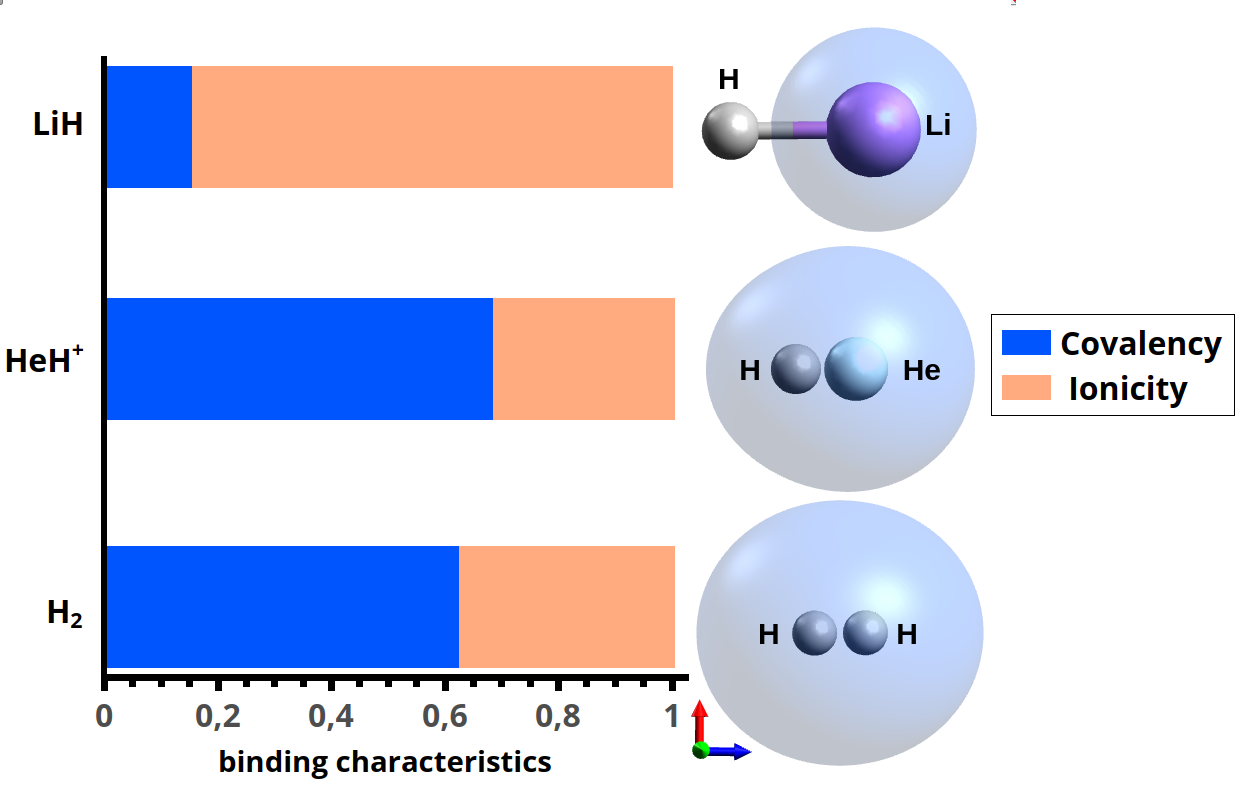}
    \caption{Illustrative picture of the molecular orbitals (isosurface probability density cut = 0.02) on the right; relative covalency and ionicity contributions, without the atomicity extracted, on the left.}
    \label{fig:covion}
\end{figure}

One can apply the same method of approach (EDABI) to other simple molecular systems. In Fig. \ref{fig:bind_heh}ab we have plotted the ground state energy \textit{versus} $R$ for \ch{HeH+} and \ch{LiH} 
and compare is with the results obtained from other methods. As one can see, the results from different 
methods converge relatively quickly, even before reaching the Mottness (shaded) asymptotic regime. This 
trend is also observed in the $R$ dependence of the size of orbitals in those two systems, as depicted in
Figs. \ref{fig:alpha_h2}ab. The $1s$ orbital size approaches very rapidly the asymptotic (atomic) 
limiting value in \ch{HeH+}, whereas the second, $2s$ electron, is responsible for the main part of the bonding
with the increasing $R$ beyond $R_{bond}$. In \ch{LiH} the situation is different as with the increasing $R$
the atomicity is enhanced and is signalling a robust formation of an almost purely ionic state. Note that the 
extraction of the atomicity requires a further analysis in this case, as we have to introduce a separate 
atomicity factor for $1s$ and $2s$ original states. A detailed analysis, as well as the results depicted in Fig. \ref{fig:covion}ab are of numerical character only, as in
this case with two different orbitals an analytic discussion is not possible (e.g., we should use the full matrix form of \eqref{eq:Hammatrix}).
To summarize, the related molecular systems such as those discussed briefly above, require a further analysis to extract the atomicity in heterogeneous systems, independently of the degree of covalency in the physical ground state. This statement is illustrated in Fig. \ref{fig:covion} with the starting degree of covalency/ionicity for those systems considered. Additionally, the values of calculated microscopic parameters of Hamiltonian \eqref{eq:full_Hamiltonian}, as well as of other microscopic characteristics, are provided in our original papers \cite{Hendzel1} and \cite{Hendzel2} (in the latter paper in the section \textit{Supporting Information}).

\subsection{Essential extension: The hydrogen bond -- an outline}
\begin{figure}[h]
    \centering
    \includegraphics[width=0.5\textwidth]{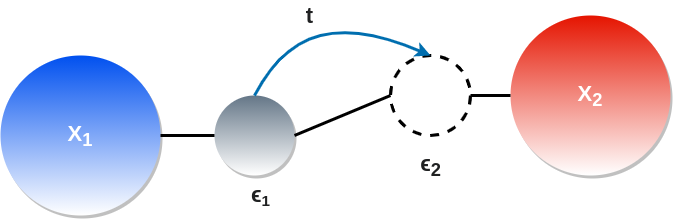}
    \caption{Schematic representation of proton resonant hopping of magnitude $t$, as the simplest model of hydrogen bond. X$_1$ and X$_2$ symbolize donor and acceptor of molecules. Different reference orbital energies $\epsilon_1$ and $\epsilon_2$ refer to two possible proton positions. }
    \centering
    \label{fig:simple_hydrogen}
\end{figure}

Recently, application of a similar model based on the second quantization representation to description of the hydrogen bond has been formulated \cite{Pulsuk1, Pulsuk2, Brovarets, Witkowski, Grabowski}. In that model the resonating fermion is proton and is taking between two more complex molecules. In particular, the question of defining covalency in that case has been raised \cite{Dereka, Lubbe, McKenzie}. Simply speaking, the proton hops between two (in)equivalent positions, in an analogical situation to that for electrons between two protons in \ch{H2} molecule. In the simplest picture of such a hydrogen bond one has the situation with quantum tunneling of a proton regarded now as a quantum particle between two positions with energy $\epsilon_1$ and $\epsilon_2$, created by two larger donor/acceptor groups $X_1$ and $X_2$ as shown in Fig. \ref{fig:simple_hydrogen}. If we ignore the explicit electron charge--density shift in $X_1$ and $X_2$, associated with the proton hopping, then the proton between the two positions labeled with $\epsilon_1$ and $\epsilon_2$ can be modeled by Hamiltonian
\begin{figure}
    \centering
    \includegraphics[width=0.8\textwidth]{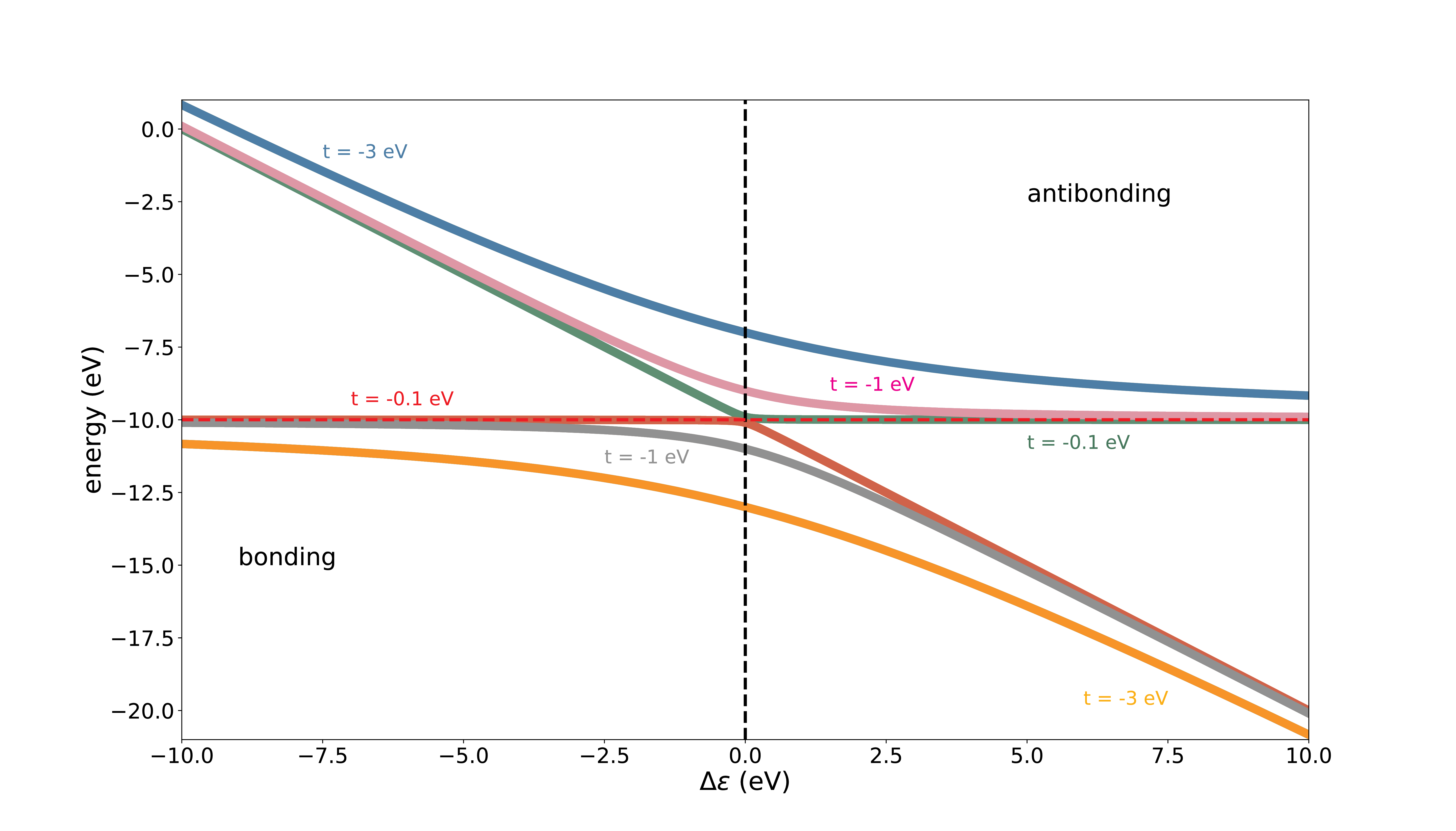}
    \caption{Bonding and antibonding states for resonating fermion (proton) between two sites as a function of the atomic--energy difference. Three values of the hopping $t$ have been selected.}
    \label{fig:res_eval}
\end{figure}

\begin{align}
    \mathcal{\hat{H}} =  \epsilon_1\hat{a}^{\dagger}_1\hat{a}_1 + \epsilon_2\hat{a}^{\dagger}_2\hat{a}_2
    + t(\hat{a}^{\dagger}_1\hat{a}_2+\hat{a}^{\dagger}_2\hat{a}_1)
\end{align}

\noindent
where $t$ is the hopping and $\hat{a}^{\dagger}_i$($\hat{a}_i$) is spinless proton creation (annihilation) operator at site $\epsilon_i$. The energy shift $|\epsilon_1-\epsilon_2| \neq 0$ appears as the proton hops between the two sites. This simple two--site model leads to bonding and antibonding states of the form

\begin{align}
    x_{1,2} = \frac{1}{2}[(\epsilon_1+\epsilon_2) \mp \sqrt{(\epsilon_1-\epsilon_2)^2 + 4t^2}]. 
\end{align}

\noindent
The eigenenergies are degenerate with respect to proton spin direction. Those energies are depicted as a function of $\Delta\epsilon \equiv \epsilon_1-\epsilon_2$ in Fig. \ref{fig:res_eval} for different $t$ values. Such a simple model provides the bonding--antibonding. The corresponding eigenfunctions can be obtained easily will not be reproduced here. 
\begin{figure}
    \centering
    \includegraphics[width=0.8\textwidth]{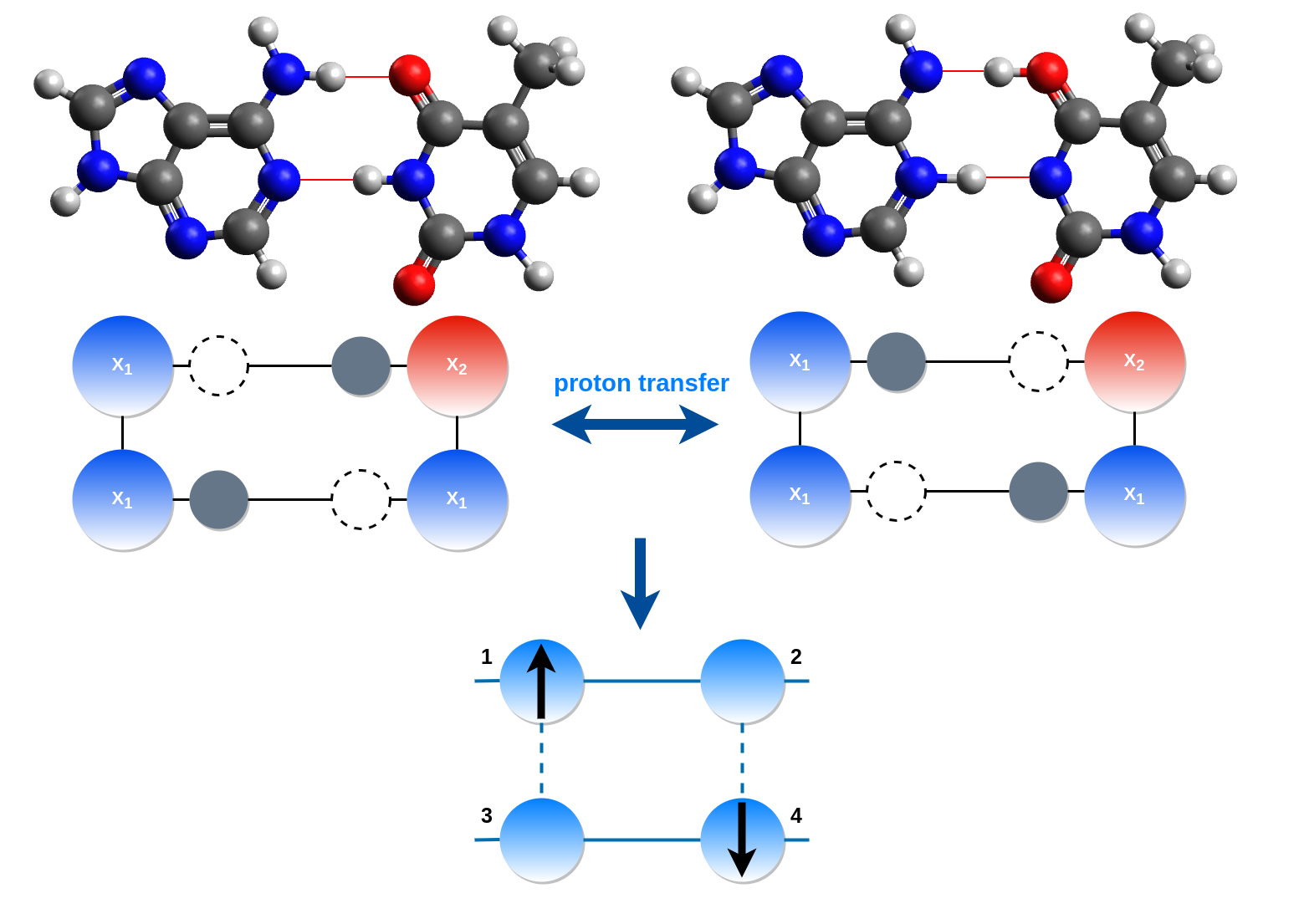}
    \caption{Top row: Two resonating proton configuration before (left) and after correlated pair hopping. Adenine (A) and tymine (T) pair has been selected. Middle row: Schematic representation of the processes depicted in the top row. Bottom row: $4$-site model with two protons with their resonating between the sites. Dashed vertical lines represent virtual hopping between the rungs. An inclusion of the electron density shift may be included by vertical hops of the electrons towards the hopping proton.}
    \label{fig:amtym}
\end{figure}

The more realistic situation is drawn in Fig. \ref{fig:amtym} for the adenine--tymine exemplary case. In this situation there are two hydrogen bonds as shown, as well as two proton transfers. The resultant schematic model is drawn at the bottom, where now both the protons spins and transverse (vertical) hopping processes $\sim t'$ have both been marked. This bottom part leads to the model with two fermions (protons) in $4$-site system. It leads to $\binom{8}{2} = 24$ spin--singlet and --triplet states. The Hamiltonian is of the form similar to expression \eqref{eq:full_Hamiltonian}, which can be analyzed formally rigorously. We omit here a detailed examination of the resultant states. This example is to show only that the model constructed for simple molecules can be readily constructed and adopted also to more involved situations. Namely, the hopping of the proton to the particular site can be associated with a motion of charge compensating electron towards the incoming proton to the closest site, i.e., the vertical electron hopping towards the proton (hopping of two electrons in total, in opposite direction, to that of each of the proton). Such a simple model requires determination of $\binom{4}{2}^4 = 1296$-fermion states. In essence, the double hydrogen bond in that case is composed of both the intersite proton resonance of a covalent type and associated with it electrostatic attraction of electron to the hopping proton in its final state. We should see a progress along this lines in near future.

\section{Outlook}

In this Chapter we have analyzed in detail the nature of covalency and have elaborated on it by defining an \textit{intrinsic} (\textit{true}) \textit{covalency}, in addition to \textit{ionicity}, as well as have defined the degree of \textit{atomicity} in nominally covalent state. With introduction of the atomicity we have defined a degree of atomic character in the resultant bound state. Our analysis starts with the exact solution of the Heitler--London model of bonding in \ch{H2} molecule as well as extends it. The analyzed solution and bonding properties are carried out explicitly and this possible by starting from the analytic solution of the model. This solution, in the closed form, is in turn, possible with the help of our original EDABI (\textbf{E}xact \textbf{D}iagonalization \textbf{Ab I}nitio) approach.

The method detailed for the \ch{H2}--molecule complementary characterization can also be applied to other simple molecular systems such as \ch{LiH} and \ch{HeH+}, each discussed briefly. Additionally, a similar modeling can also be carried out for the analysis of hydrogen bond, as briefly outlined at the end. We should be able to see an important progress in understanding these systems along these lines in the near future. 

\section{Acknowledgement}

This work was supported by Grants OPUS No.~UMO-2018/29/B/ST3/02646 and No.~UMO-2021/41/B/ST3/04070 from Narodowe Centrum Nauki. We would like to thank our colleagues from the Chemistry Department of the Jagiellonian University for numerous
discussions and general remarks during the course of this project.

\thispagestyle{empty}

\vspace*{4mm}

\begin{center}
KEYWORDS
\end{center}
\noindent hydrogen (\ch{H2}) molecule, atomicity in molecular states, Exact solution of Heitler--London model, Exact diagonalization \textit{ab initio} (EDABI) method

\end{document}